\documentclass[conference,compsoc]{IEEEtran}
\usepackage{graphicx}
\usepackage{subfigure}
\usepackage{float}
\usepackage{verbatim}
\usepackage{multirow}

\ifCLASSOPTIONcompsoc
  \usepackage[nocompress]{cite}
\else
  \usepackage{cite}
\fi
\ifCLASSINFOpdf
\else
\fi

\usepackage{xcolor}
\usepackage{xspace}
\usepackage{hyperref}

\makeatletter
\newcommand{\linebreakand}{%
  \end{@IEEEauthorhalign}
  \hfill\mbox{}\par
  \mbox{}\hfill\begin{@IEEEauthorhalign}
}
\makeatother

\begin{document}
%

%
\title{Autotuning PolyBench Benchmarks with LLVM Clang/Polly Loop Optimization Pragmas Using Bayesian Optimization}
%


\author{\IEEEauthorblockN{Xingfu Wu, Michael Kruse, Prasanna Balaprakash, Hal Finkel, Paul Hovland, Valerie Taylor}
\IEEEauthorblockA{Argonne National Laboratory, Lemont, IL 60439 \\
Email: \{xingfu.wu,michael.kruse,pbalapra,hfinkel,hovland,vtaylor\}@anl.gov}
\and
\linebreakand 
\IEEEauthorblockN{Mary Hall}
\IEEEauthorblockA{University of Utah, Salt Lake City, UT 84103 \\ 
Email: mhall@cs.utah.edu}
}


\maketitle

\begin{abstract}

An autotuning is an approach that explores a search space of possible implementations/configurations of a kernel or an application by selecting and evaluating a subset  of implementations/configurations on a target platform and/or use models to identify a high performance implementation/configuration. In this paper, we develop an autotuning framework that leverages Bayesian optimization to explore the parameter space search. We select six of the most complex benchmarks from the application domains of the PolyBench benchmarks (syr2k, 3mm, heat-3d, lu, covariance, and Floyd-Warshall) and apply the newly developed LLVM Clang/Polly loop optimization pragmas to the benchmarks to optimize them. We then use the autotuning framework to optimize the pragma parameters to improve their performance. The experimental results show that our autotuning approach outperforms the other compiling methods to provide the smallest execution time for the benchmarks syr2k, 3mm, heat-3d, lu, and covariance with two large datasets in 200 code evaluations for effectively searching the parameter spaces with up to 170,368 different configurations. We compare four different supervised learning methods within Bayesian optimization and evaluate their effectiveness. We find that the Floyd-Warshall benchmark did not benefit from autotuning because Polly uses heuristics to optimize the benchmark to make it run much slower. To cope with this issue, we provide some compiler option solutions to improve the performance.

\end{abstract}



\section{Introduction}

As the complexity of high performance computing (HPC)  ecosystems (hardware stacks, software stacks, applications) continues to rise, achieving optimal performance becomes a challenge. The number of tunable parameters an HPC user can configure has increased, resulting in the overall parameter space growing significantly. Exhaustively evaluating all parameter combinations becomes very time-consuming. 
Therefore, autotuning for automatic exploration of parameter space is desirable.
An autotuning is an approach that explores a search space of possible implementations/configurations of a kernel or an application by selecting and evaluating a subset of implementations/configurations on a target platform and/or use models to identify the high performance  implementation/configuration within a given computational budget. 
In this work, we develop a machine learning (ML)-based autotuning framework to reduce the parameter search space in order to autotune loop optimization pragmas to improve performance.

A large amount of literature on autotuning exists. Balaprakash et al. \cite{BD18} surveyed  the state of the practice in incorporating autotuned code into HPC applications; the authors highlighted insights from prior work and identified the challenges in advancing autotuning into wider and long-term use. Traditional autotuning methods are built on heuristics that derive from experience \cite{TC02, CH06, GO10} and model-based methods \cite{TH11, BG13, FE17}. At the compiler level, machine-learning-based methods are used for automatic tuning of the iterative compilation process \cite{OP17} and tuning of compiler-generated code \cite{TC09, MS14}. Some recent work has used machine learning and sophisticated statistical learning methods to reduce the overhead of autotuning \cite{RB16, MA17, TN18}. Most recent work on autotuning of OpenMP code has gone beyond loop schedules to look at parallel tasks and function inlining \cite{SJ19, KC14, MA11}. In particular, a lightweight framework was proposed to enable autotuning of OpenMP pragmas to ease the performance tuning of OpenMP codes across platforms \cite{SJ19}; the approach incorporated the Search using Random Forests (SuRF). 
SuRF is the earliest version of the parameter space search ytopt \cite{YTO}. Currently, ytopt leverages Bayesian optimization \cite{BH19,SL12} to explore the parameter space search and uses different supervised learning methods within Bayesian optimization such as random forests, Gaussian process regression, extra trees, and gradient boosted regression trees.

Kruse and Finkel \cite{KF19} implemented a newly proposed prototype of user-directed loop transformations using Clang and Polly \cite{POLL} with additional loop transformation pragmas such as loop reversal, loop interchange, tiling, and array packing in DOE's Exascale Computing Project (ECP) SOLLVE  \cite{SOLL}. SOLLVE seeks to deliver a high-quality, robust implementation of OpenMP and project extensions in LLVM \cite{LLVM}, which is a collection of modular and reusable compiler and toolchain technologies. Research is needed to determine how to efficiently combine these loop transformation pragmas to optimize an application. Because of the large parameter space of these pragmas and related parameters, autotuning for automatic exploration of the parameter space is desirable. In our preliminary work with the Y-TUNE project \cite{WK20}, we worked on integrating the loop optimization pragmas within the ytopt package to autotune loop optimization pragmas for optimal performance. In this paper, we develop an autotuning framework to integrate the Clang/Polly loop optimization pragmas with the ytopt, and we apply the loop optimization pragmas to the PolyBench benchmarks \cite{YP16} to evaluate the autotuning framework.

PolyBench 4.2 \cite{YP16} is a benchmark suite of 30 numerical computations extracted from operations in various application domains (linear algebra computations, image processing, physics simulation, and data mining). In this work, we select six of the most complex benchmarks from the application domains of PolyBench benchmarks (syr2k, 3mm, heat-3d, lu, covariance, and Floyd-Warshall) and apply the newly developed LLVM Clang/Polly loop optimization pragmas to these benchmarks to improve their performance. 

We evaluate the performance on a machine with 3.1 GHz Quad-core Intel Core i7 and 16 GB of memory. The experimental results show that the autotuning outperforms the other compiling methods to provide the smallest execution time for the benchmarks syr2k, 3mm, heat-3d, lu, and covariance with two datasets in 200 evaluations for effectively searching the parameter spaces with up to 170,368 different configurations. We compare four different supervised ML methods within Bayesian optimization and evaluate their effectiveness. We find that one exception for Polly is the Floyd-Warshall benchmark because Polly uses heuristics to optimize the benchmark to make it run much slower. To cope with this issue, we provide some compiler option solutions to improve the performance.

This paper makes the following contributions:
 \begin{itemize}
\item We develop an autotuning framework to leverage Bayesian optimization to explore the parameter space search with the newly developed Clang loop optimization pragmas.
\item We apply the loop optimization pragmas to the PolyBench benchmarks to optimize them.
\item We show that the autotuning framework outperforms other compiling methods to achieve the optimal implementation in 200 code evaluations for effectively searching the parameter spaces with up to 170,368 different configurations.
\item We compare four different supervised learning methods within Bayesian optimization and evaluate their effectiveness. 
\end{itemize}

The remainder of this paper is organized as follows. Section 2 discusses SOLLVE Clang/Polly loop optimization pragmas and the parameter space search ytopt and then presents an autotuning framework based on them. Section 3 surveys the PolyBench benchmarks and selects six of the most complex benchmarks from the application domains. Section 4 applies the Clang loop optimization pragmas to these benchmarks to improve them and then use the autotuning framework to autotune the pragma parameters to achieve the optimal performance. We also compare four different supervised learning methods within Bayesian optimization and evaluate their effectiveness. Section 5 summarizes our conclusions and briefly discusses future work. 


\section{An Autotuning Framework}

In this section, we discuss loop optimization pragmas implemented in LLVM Clang/Polly and the ytopt parameter search space autotuner. We then present an autotuning framework based on them.

\subsection{Clang/Polly Loop Optimization Pragmas}

Compiler directives such as pragmas can help programmers to separate an algorithm's semantics from its optimization. Pragma directives for code transformations are useful for assisting program optimization and are already widely used in OpenMP. In~\cite{KF19}, a prototype of user-directed loop transformations using Clang and Polly~\cite{POLL} was implemented for the US DoE's ECP SOLLVE project~\cite{SOLL}.
Polly is LLVM's polyhedral loop optimizer which makes it easy to apply specific transformations as directed by pragmas.
We used the SOLLVE project's development branch for LLVM located at \url{https://github.com/SOLLVE/llvm-project/tree/pragma-clang-loop}.
While the SOLLVE team is working on integrating the changes into the official LLVM repository, only few of the changes have been upstreamed yet.
The additional loop transformation directives supported are loop reversal (inverting the iteration order of a loop), loop interchange (permuating the order of nested loops), tiling, unroll(-and-jam), array packing (temporarily copying the data of a loop's working set into a new buffer) and thread parallelization. More importantly, it supports composing multiple loop nest transformation in arbitrary order.
Vectorization is also supported by LLVM's dedicated loop vectorizer.
These pragmas are intended to make applying common loop optimization technique much easier and allow better separation of a code's semantics and its optimization. In this paper, we use some of these pragmas to optimize several PolyBench benchmarks and then propose the framework to autotune them.

\begin{figure}
\center
 \includegraphics[width=.45\textwidth]{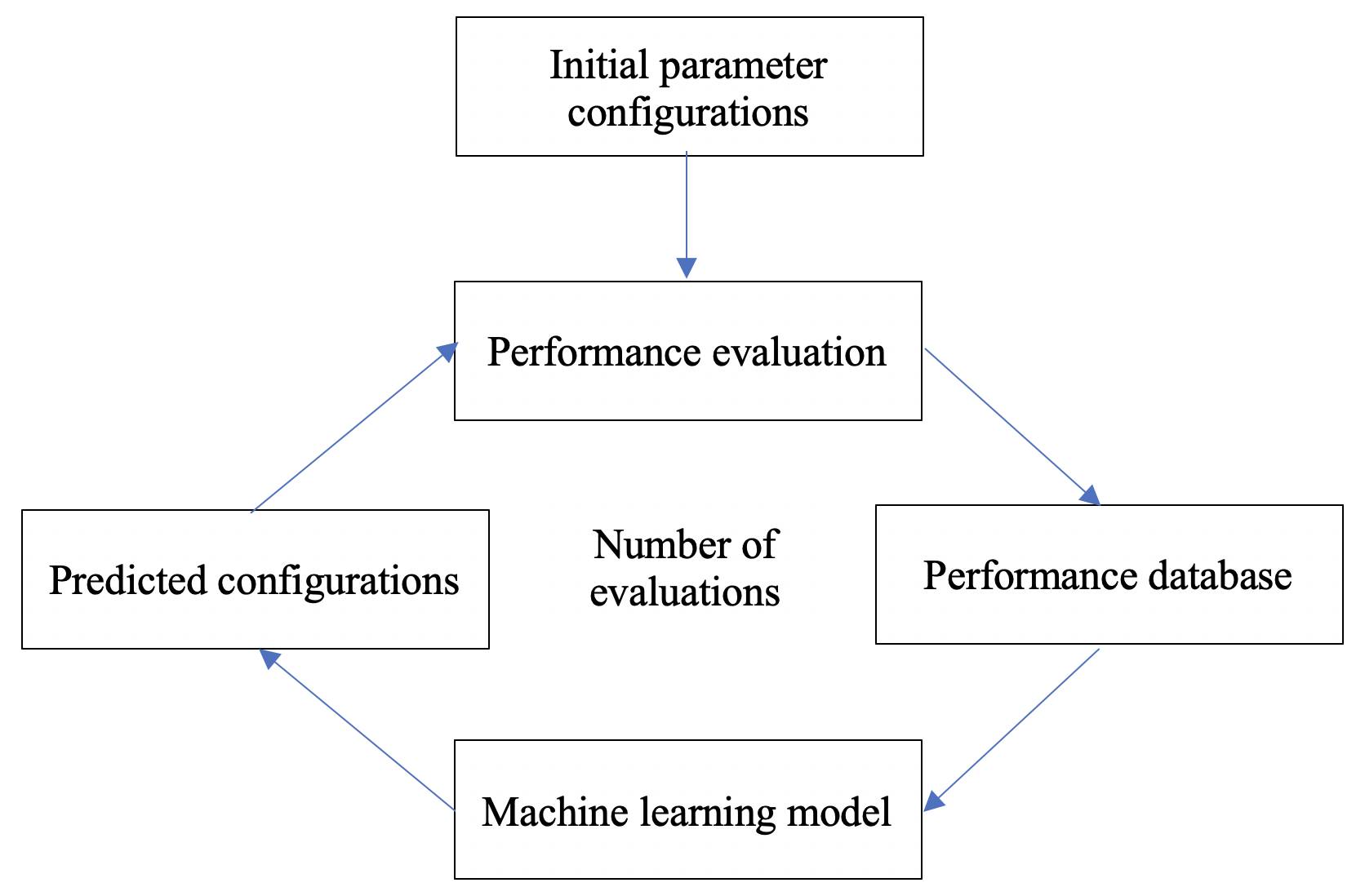}
 \caption{Machine-learning-based autotuner ytopt}
\label{fig1}
\end{figure}

\subsection{Parameter Space Search}

ytopt \cite{YTO} is a machine-learning-based parameter space search software package that leverages Bayesian optimization \cite{BH19,SL12} to explore the parameter space search and uses different supervised ML methods within Bayesian optimization such as random forests, Gaussian process regression, extra trees, and gradient boosted regression trees. A high level overview of the package is shown in Figure \ref{fig1}. The package takes the user-defined parameter space definition (bounds and constraints) and the parameter configuration evaluation interface as input. The initialization phase consists  of  sampling  a  small  number  of  input  parameter  configurations through random sampling or Latin hypercube sampling and recording the performance to a performance database (performance output files include the configuration, runtime, and wall-clock time). The input-output pairs are used to fit a surrogate model using a ML method. The iterative phase of search consists in sampling an input parameter configuration for evaluation by progressively leveraging and refining the surrogate model. At the evaluation stage, check the performance database to make sure that this chosen configuration is new. If it was evaluated before, skip the evaluation.
To that end, the search uses Bayesian optimization in which uncertainty quantification of the surrogate model is leveraged to balance exploration of the search space and identification of more-promising regions using lower confidence bound acquisition function. 

Three of the four ML methods---random forest, Extra Trees, and gradient boosted regression trees---follow the autotuning process based on the performance database shown in Figure \ref{fig1}. Gaussian process, however, still uses random or Latin hypercube sampling to generate the parameter configurations for performance evaluation.

ytopt is a Python package that uses scikit-optimize \cite{SCIO}, autotune \cite{AUTO}, and ytopt subpackage \cite{YTO}. See our initial work \cite{WK20} for the detailed installation and download information. It uses ConfigSpace \cite{CFS} package to handle the algebraic constraints on the parameter configuration space. 

\subsection{Proposed Autotuning Framework}

Based on the Clang loop optimization pragmas and the parameter space search ytopt, we present the general autotuning framework in the following steps shown in Figure \ref{fig2}:

\begin{figure}
\center
 \includegraphics[width=.47\textwidth]{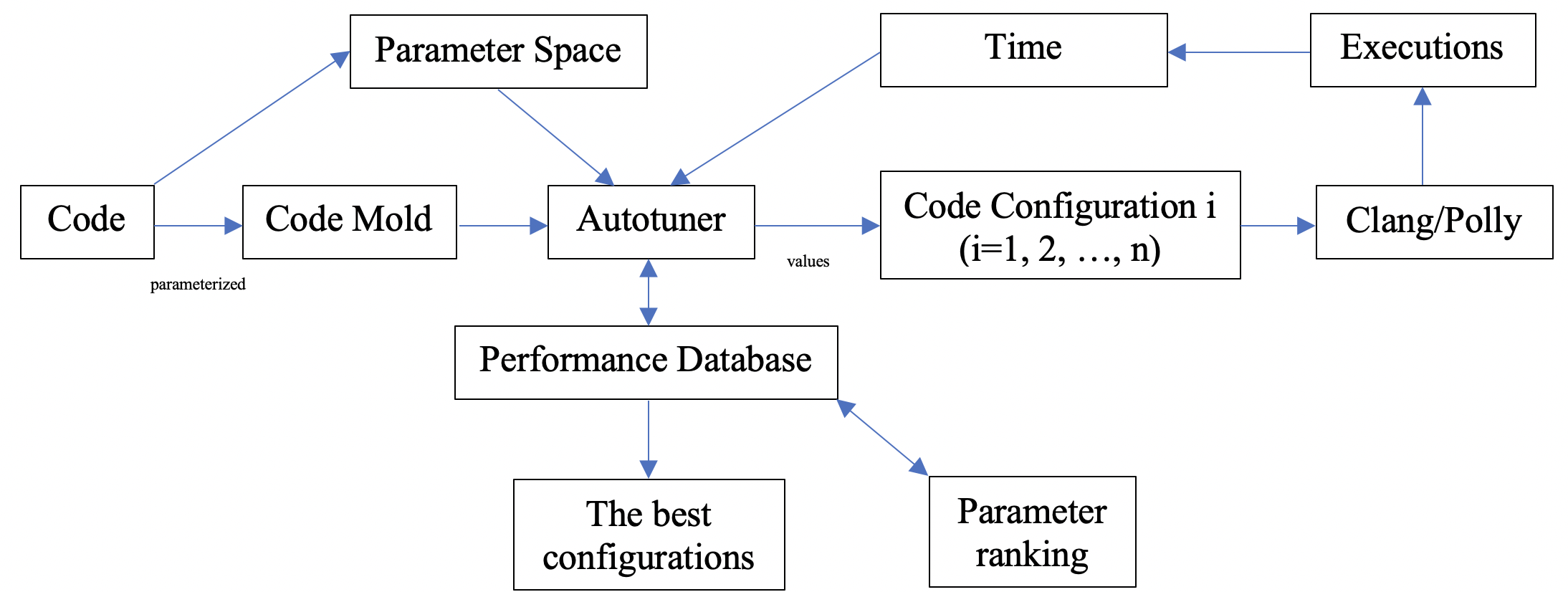}
 \caption{Framework for autotuning Clang/Polly Loop pragmas}
\label{fig2}
\end{figure}

 \begin{itemize}
\item [1)] Analyze an application code to identify the important parameters that we try to focus on.
\item	[2)] Replace these parameters with  symbols such as \#P0, \#P1, \#P2, ..., \#Pm in  top-down order to generate another code with these symbols as a code mold. 
\item	[3)] Define the value ranges of these symbols for the parameter space as an input of the ytopt autotuner (problem.py).
\item	[4)] Use the ytopt autotuner to search the parameter space to assign the values in the allowed ranges (using random forest as default), and replace these symbols in the mold code with them to generate a new code using the function plopper (plopper.py).
\item	[5)] Use the plopper to compile the code and execute it to get the execution time (using a Perl script exe.pl).
\item	[6)] Use the autotuner to write the execution time and the elapsed time with the parameters' values to the performance database (two output files: results.csv and results.json).
\item	 [7)] Repeat steps 4, 5, and 6 until  reaching the maximum number of code evaluations n (using the option --max-evals=maximum number n; default: 100).
\item	[8)] Process the database to find the smallest execution time and output the optimal configuration for the execution time (findMin.py)
\item [9)] Identify the most important features which impact the performance for the search improvement.
 \end{itemize}

This autotuning framework requires the following components: configspace, scikit-optimize, autotune, ytopt, and LLVM clang/polly. The framework provides the following main options: 
 \begin{itemize}
\item[]  --max-evals: to set the maximum number of evaluations n
\item[]  --learner: to set the ML method using random forests (RF),  Extra Trees (ET), gradient Boosted regression trees (GBRT), or Gaussian processes (GP). The default is RF.
\end{itemize}

In this work, we apply the Clang loop optimization pragmas to the PolyBench benchmarks \cite{YP16} to evaluate the autotuning framework using the four different ML methods.

\section{PolyBench Benchmarks}

PolyBench 4.2 \cite{YP16} is a benchmark suite of 30 numerical computations extracted from operations in various application domains (19 linear algebra computations, 3 image-processing applications, 6 physics simulations, and 2 data-mining applications). The details about the benchmarks are as follows:
 \begin{itemize}
\item	[1)] Linear Algebra: 

BLAS (7):  gemm,	gemver,	 gesummv,	symm,	\textbf{\color{red} syr2k},	syrk,	trmm

Kernels (6): 2mm,	\textbf{\color{red} 3mm},	atax,	bicg,	doitgen, 	mvt

Solvers (6): Cholesky,  durbin,  gramschmidt,	 \textbf{\color{red} lu},	ludcmp,		trisolv

\item	[2)] Medley (image processing) (3): deriche,	\textbf{\color{red} floyd-warshall},	  nussinov

\item	[3)] Physics simulation (stencils) (6): adi,  fdtd-2d,  \textbf{\color{red} heat-3d},  jacobi-1d,  jacobi-2d,  seidel-2d

\item	[4)] Data mining (2): correlation,  \textbf{\color{red} covariance}
\end{itemize}

In this work, we choose the most complex benchmark with the most levels of nested loops (the red color) from each group to illustrate how the autotuning framework performs, and we compare their performance.

syr2k is a symmetric rank 2k update from BLAS and entails the matrix multiplication C = A*alpha*B+ B*alpha*A+belta*C, where A is an NxM matrix, B is an MxN matrix, and C is an NxN symmetric matrix. We use the following large datasets: LARGE\_DATASET (M 1000, N 1200) and EXTRALARGE\_DATASET (M 2000, N 2600) for our case study.

3mm is one of the linear algebra kernels that consists of three matrix multiplications and entails G=(A*B)*(C*D), where Ais a PxQ matrix; B is a QxR matrix; C is an RxS matrix; and D is an SxT matrix. We use the following large datasets: LARGE\_DATASET (P 800, Q 900, R 1000, S 1100, T 1200) and EXTRALARGE\_DATASET (P 1600, Q 1800, R 2000, S 2200, T 2400).

lu is LU decomposition without pivoting in linear algebra solvers and entails A = L*U, where L is an NxN lower triangular matrix and U is an NxN upper triangular matrix. We use the following large datasets : LARGE\_DATASET (N 2000) and EXTRALARGE\_DATASET (N 4000).

heat-3d entails a heat equation over 3D space in Stencil. Stencil computations iteratively update a grid of data using the same pattern of computation. We use the following large datasets: LARGE\_DATASET (TSTEPS 500, N 120) and EXTRALARGE\_DATASET (TSTEPS 1000, N 200).

covariance entails computing the covariance, a measure from statistics that shows how linearly related two variables are. It takes the data (NxM matrix that represents N data points, each with M attributes) as input and gives the cov (symmetric MxM matrix where the i,jth element is the covariance between i and j) as the output. We use the following large datasets: LARGE\_DATASET (M 1200, N 1400) and EXTRALARGE\_DATASET (M 2600, N 3000).

Floyd-Warshall entails computing the shortest paths between each pair of nodes in a graph in Medley. The input is an NxN matrix, where the i,jth entry represents the cost of taking an edge from i to j. The output is an NxN matrix, where the i,jth entry represents the shortest path length from i to j.  We use the following datasets: MEDIUM\_DATASET (N 500) and LARGE\_DATASET (N 2800).

\section{Autotuning the Benchmarks with Clang Loop Optimization Pragmas}

We apply the Clang loop optimization pragmas \cite{KF19} to the chosen PolyBench benchmarks to optimize them. We define the parameters for these pragmas and then use the framework to autotune the pragma parameters to achieve the optimal performance. Based on the performance database, we write the Python script findMin.py to find the smallest execution time and output the best configurations. We also use four different ML methods to investigate how they impact finding the optimal configuration from the input parameter search space based on the performance database and what the optimal loop tiling sizes. We evaluate the performance on a machine with 3.1 GHz Quad-core Intel Core i7 and 16 GB of memory and 1 TB SSD; gcc 7.2 and clang 10.0 are installed on the machine.

\subsection{Case Study: syr2k with multiple loop transformations (loop tiling, interchange, and array packing)}

We apply multiple loop transformations such as tiling, interchange, and array packing pragmas to the benchmark syr2k for this case study. We assume that the loop tiling (\#pragma clang loop(i,j,k) tile sizes( )) is already applied to syr2k. We define the following parameters:

{\scriptsize
\begin{verbatim}
#P0
#P1
#P2
#pragma clang loop(i,j,k) tile sizes(#P3,#P4,#P5) 
	   		floor_ids(i1,j1,k1) tile_ids(i2,j2,k2)
#pragma clang loop id(i)
  for (i = 0; i < _PB_N; i++) {
    #pragma clang loop id(j)
    for (j = 0; j < _PB_M; j++) {
     #pragma clang loop id(k)
        for (k = 0; k <= i; k++)
        {
          C[i][k] += A[k][j]*alpha*B[i][j] 
          			+ B[k][j]*alpha*A[i][j];
        }
    }
  }
\end{verbatim}
 }  
 
Based on these parameters, we create a code mold. For this case, we have to make sure that there is no dependence among P0, P1, P2, P3, P4, and P5. Based on the defined six parameters, we have the following parameter space input\_space using ConfigSpace:

{\scriptsize
\begin{verbatim}
cs  CS.ConfigurationSpace(seed=1234)
P0=CSH.CategoricalHyperparameter(name='P0', 
choices=["#pragma clang loop(j2) pack array(A) 
allocate(malloc)", " "], default_value=' ')
P1=CSH.CategoricalHyperparameter(name='P1', 
choices=["#pragma clang loop(i1) pack array(B) 
allocate(malloc)", " "], default_value=' ')
P2=CSH.CategoricalHyperparameter(name='P2', 
choices=["#pragma clang loop(i1,j1,k1,i2,j2) 
interchange permutation(j1,k1,i1,j2,i2)", " "], default_value=' ')
P3=CSH.OrdinalHyperparameter(name='P3', 
sequence=['4','8','16','20','32','50','64','80','96','100',
'128'], default_value='96')
P4=CSH.OrdinalHyperparameter(name='P4', 
sequence=['4','8','16','20','32','50','64','80','100','128',
'2048'], default_value='2048')
P5=CSH.OrdinalHyperparameter(name='P5', 
sequence=['4','8','16','20','32','50','64','80','100','128',
'256'], default_value='256')
cs.add_hyperparameters([P0, P1, P2, p3, P4, P5])
cond1 = CS.InCondition(P1, P0, 
['#pragma clang loop(j2) pack array(A) allocate(malloc)'])
cs.add_condition(cond1)
input_space = cs 
\end{verbatim}
 }
 
where  the parameters P0, P1, and P2 have the choices of the pragmas or nothing. P0 is the array packing for A (\#pragma clang loop(j2) pack array(A) allocate(malloc)). P2 permutes the order of nested loops and has no dependence with P0 and P1. For P0 and P1, we add the conditions (CS.InCondition) so that Packing arrays A and B occurs at the same time.
The parameters P3, P4, and P5 represent the tile size for each loop. Based on the tile size settings in \cite{LI16, BU20} related to the cache sizes, we set the default size to 96 for P3, 2048 for P4, and 256 for P5. For simplicity, we set 11 tile sizes for these parameters. The parameter space consists of 2x2x2x11x11x11= 10,648 different configurations. Then we use our autotuning framework to find out which configuration results in the smallest execution time.

We use four ML methods---RF, ET, GBRT, and GP---to investigate which one generates the smallest runtime for which configuration in 200 evaluations for syr2k with the large dataset. We then use the best method to autotune the benchmark with the extra large dataset. 

\begin{figure}
\center
 \includegraphics[width=.45\textwidth]{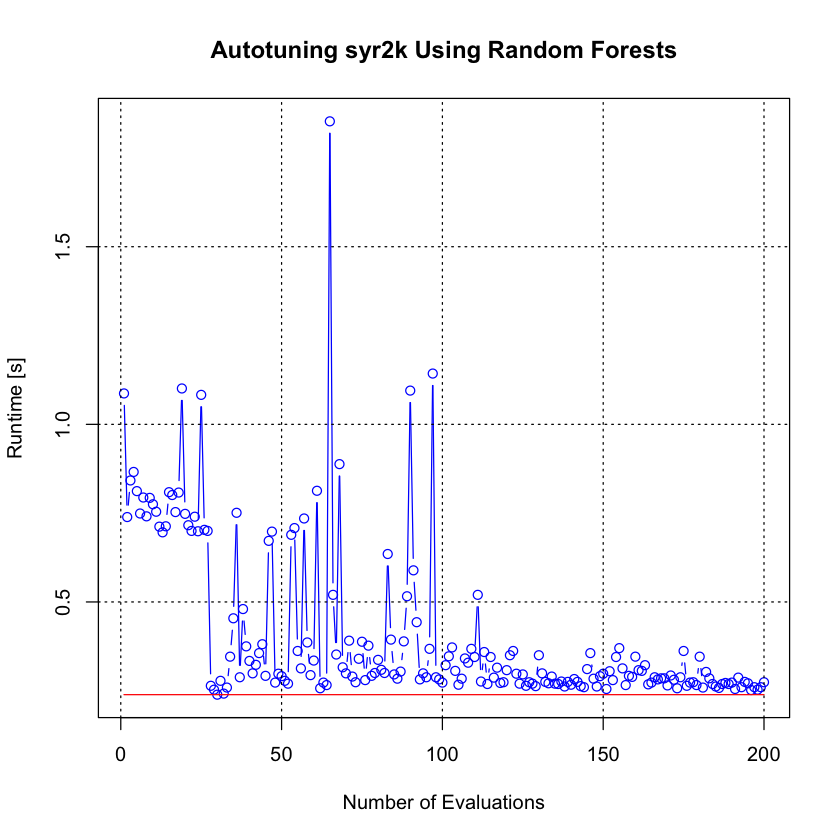}
 \caption{Autotuning syr2k using RF in 200 evaluations}
\label{fig3}
\end{figure}

In Figure \ref{fig3}, RF results in the smallest runtime of 0.239s for the configuration ('\#pragma clang loop(j2) pack array(A) allocate(malloc)','\#pragma clang loop(i1) pack array(B) allocate(malloc)','\#pragma clang loop(i1,j1,k1,i2,j2) interchange permutation(j1,k1,i1,j2,i2)',128,128,100 ) at Evaluation 30 of 200 evaluations. The blue line is for all evaluations; the red line is the best execution time. RF results in the execution time close to the smallest one with increasing the number of evaluations.

\begin{figure}
\center
 \includegraphics[width=.45\textwidth]{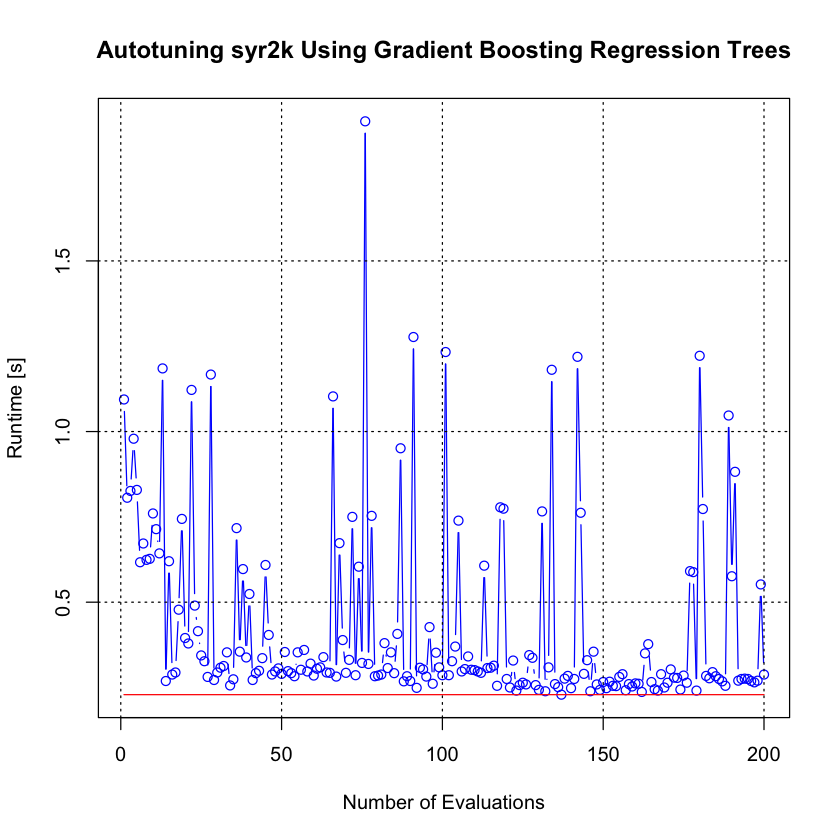}
 \caption{Autotuning syr2k using GBRT in 200 evaluations}
\label{fig4}
\end{figure}

In Figure \ref{fig4}, GBRT results in the smallest runtime of 0.229s for the configuration ('\#pragma clang loop(j2) pack array(A) allocate(malloc)','\#pragma clang loop(i1) pack array(B) allocate(malloc)','\#pragma clang loop(i1,j1,k1,i2,j2) interchange permutation(j1,k1,i1,j2,i2)',50,128,256) at Evaluation 137 of 200 evaluations. 

\begin{figure}
\center
 \includegraphics[width=.45\textwidth]{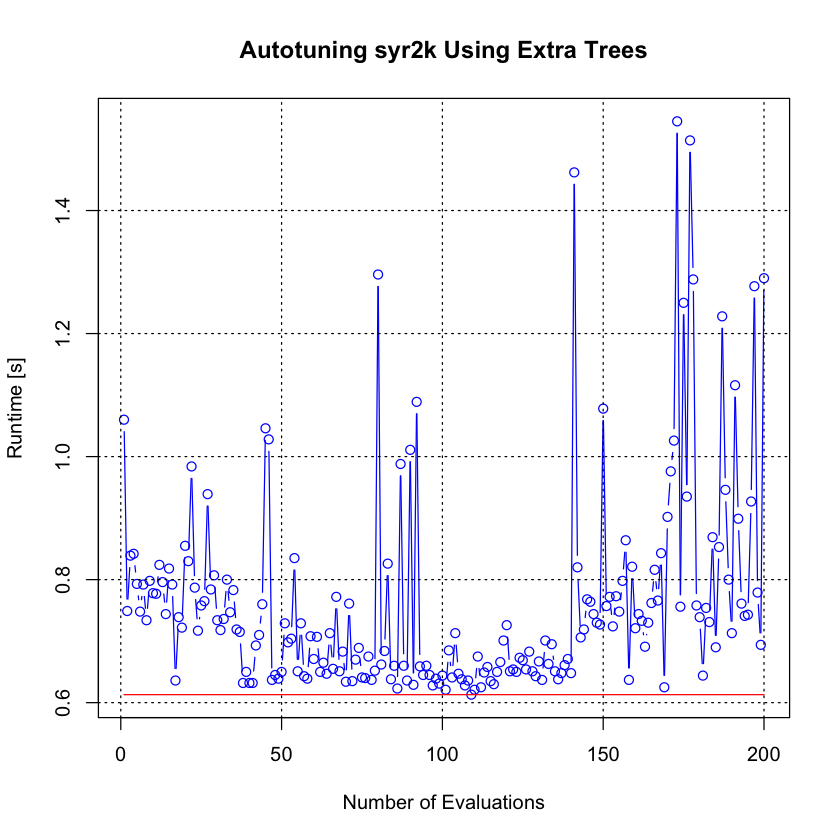}
 \caption{Autotuning syr2k using ET in 200 evaluations}
\label{fig5}
\end{figure}

In Figure \ref{fig5}, ET results in the smallest runtime of 0.613s for the configuration (' ', ' ', ' ' ,100,8,8 ) at Evaluation 109 of 200 evaluations.  ET does not show any pragmas in the best configuration with the only tile size (100, 8, 8).

\begin{figure}
\center
 \includegraphics[width=.45\textwidth]{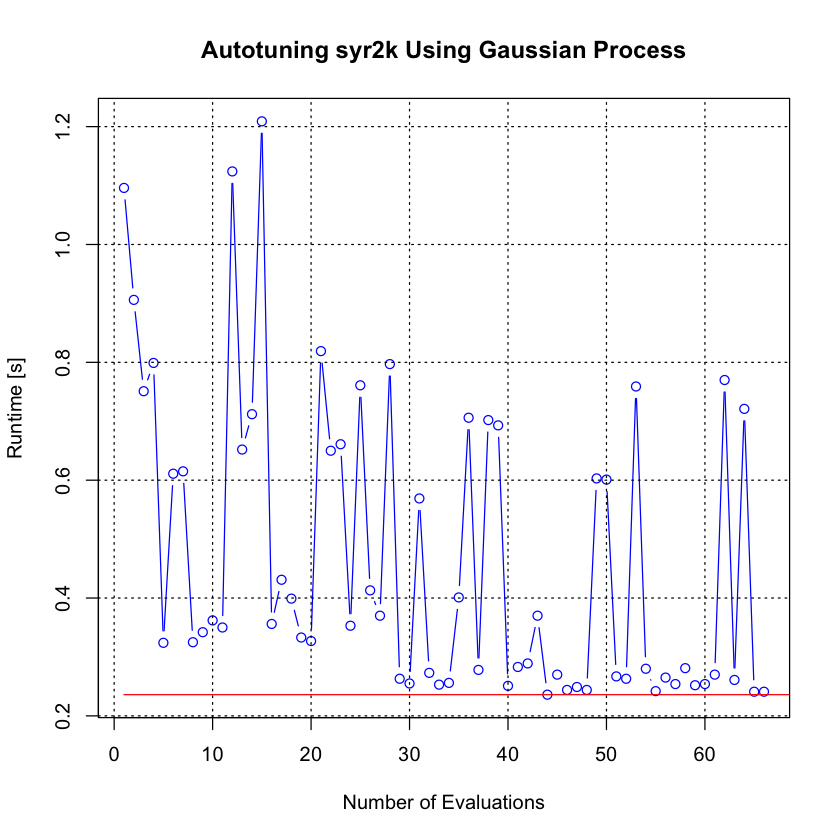}
 \caption{Autotuning syr2k using GP in 200 evaluations}
\label{fig6}
\end{figure}

In Figure \ref{fig6}, GP results in the smallest runtime of 0.236s for the configuration ('\#pragma clang loop(j2) pack array(A) allocate(malloc)','\#pragma clang loop(i1) pack array(B) allocate(malloc)','\#pragma clang loop(i1,j1,k1,i2,j2) interchange permutation(j1,k1,i1,j2,i2)',80,100,256 ) at Evaluation 44 and finishes only 66 evaluations. As  discussed in Section 2.2, GP does not use the performance database to assist the parameter space search as designed, and thus it uses only 66 of the 200 evaluations. The other 134 evaluations are skipped because of the replicated evaluations. The other methods RF, GBRT, and ET use the performance database to assist the parameter space search, and they finish all 200 evaluations with different configurations.

\begin{table}
\center
\caption{Performance (in seconds) comparison of syr2k using different compilers and autotuning}
\begin{tabular}{c}
  \includegraphics[width=.47\textwidth]{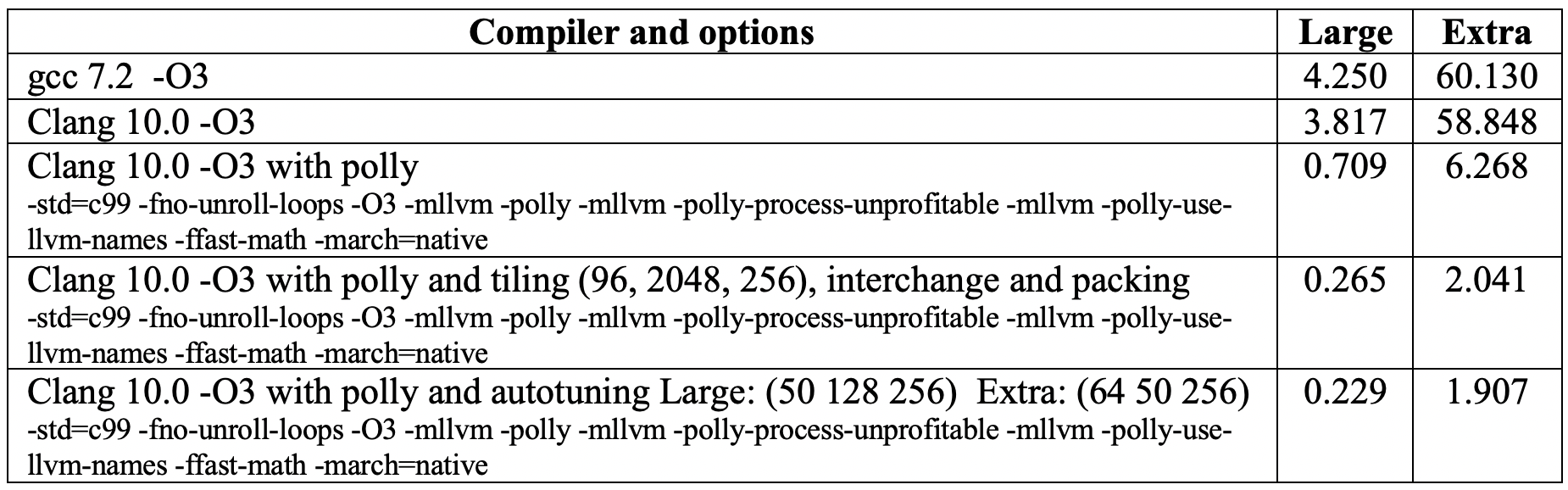}
  \end{tabular}
\label{tab1}       
\end{table} 

Table \ref{tab1} shows the performance comparison of syr2k using different compilers and autotuning. For the first three rows, these compilers and options are applied to the original code syr2k (without any loop pragma) to measure the smallest execution time in 10  runs. The fourth rows are for the syr2k with the loop tiling and default tile size (96, 2048, 256), interchange, and array packing. The last row is the results using autotuning. The best configuration in 200 evaluations has the tile size (50 128 256) for the large dataset and the tile size (64 50 256) for the extra large dataset using GBRT. We observe that autotuning outperforms the other compiling methods to provide the smallest execution time for both datasets. 

\subsection{Case Study: 3mm with multiple loop transformations} 
\begin{figure}
\center
 \includegraphics[width=.45\textwidth]{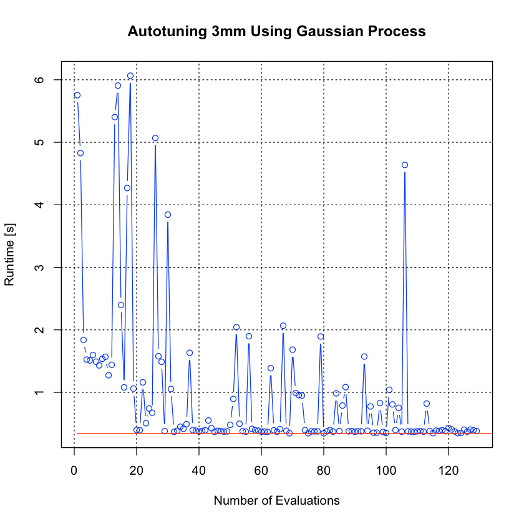}
 \caption{Autotuning 3mm using GP in 200 evaluations}
\label{fig7}
\end{figure}

We apply multiple loop transformations such as tiling, interchange, and array packing pragmas to the benchmark 3mm for this case study. We define 10 pragmas parameters to autotune the benchmark with the parameter space of 170,368 different configurations and use four ML methods---RF, ET, GBRT, and GP---to investigate which one generates the smallest runtime for which configuration in 200 evaluations for 3mm with the large dataset. We find that GP results in the smallest runtime of 0.345 s for the configuration (' ', ' ', ' ', 80, 100, 4, ' ', ' ', ' ' ,' ') at Evaluation 80 and finishes 129 evaluations as shown in Figure \ref{fig7}. Then we use GP to autotune the benchmark for the extra large dataset.

\begin{table}
\center
\caption{Performance (in seconds) comparison of 3mm using different compilers and autotuning}
\begin{tabular}{c}
  \includegraphics[width=.47\textwidth]{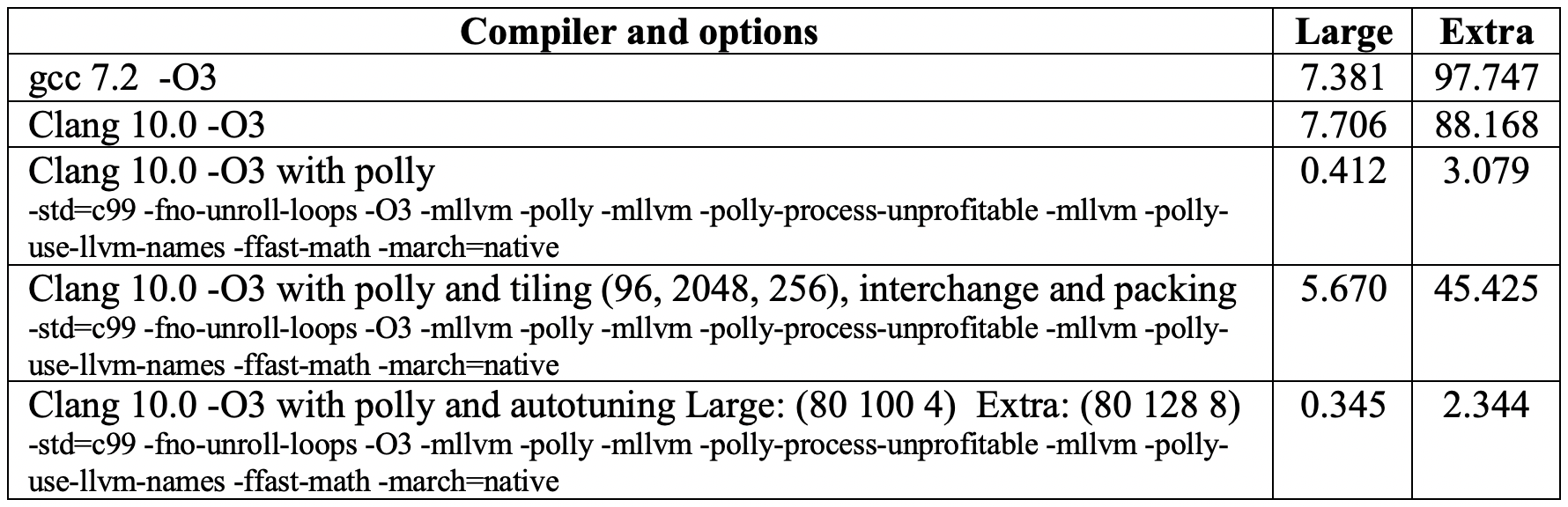}
  \end{tabular}
\label{tab2}       
\end{table} 

Table \ref{tab2} shows the performance comparison of 3mm using different compilers and autotuning. For the first three rows, these compilers and options are applied to the original code 3mm (without any loop pragma) to measure the smallest execution time in 10 same runs. The fourth rows are for the 3mm with the loop tiling with default tile size (96, 2048, 256), interchange, and array packing. The last row is the results using autotuning. The best configuration in 200 evaluations has the tile size (80 100 4)  for the large dataset and the tile size (80 128 8) for the extra large dataset using GP. We observe that autotuning outperforms the other methods to provide the smallest execution time for both datasets. 

\subsection{Case Study: lu with multiple loop transformations}

\begin{figure}
\center
 \includegraphics[width=.45\textwidth]{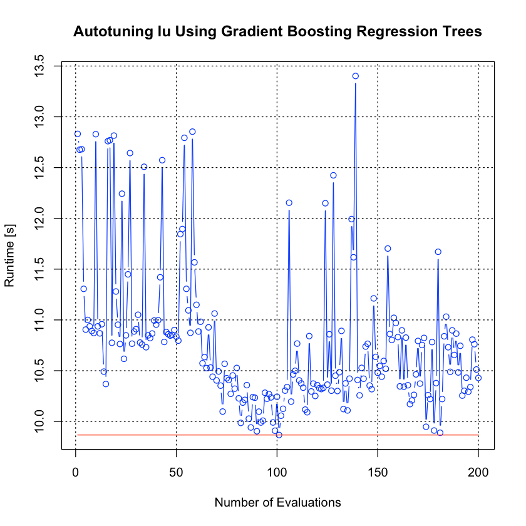}
 \caption{Autotuning lu using GBRT in 200 evaluations}
\label{fig8}
\end{figure}

We apply multiple loop transformations such as tiling, interchange, and array packing pragmas to the benchmark lu for this case study. We define the five pragma parameters to autotune the benchmark. We use four ML methods---RF, ET, GBRT, and GP---to check which one generates the smallest runtime for which configuration in 200 evaluations for lu with the large dataset. We find that GBRT has the smallest runtime of 9.867 s for the configuration ('\#pragma clang loop(i1) pack array(A) allocate(malloc)', ' ', 50, 2048, 8) at Evaluation 101 of 200 evaluations, as shown in Figure \ref{fig8}. Then we use GBRT to autotune the benchmark with the extra large dataset.

\begin{table}
\center
\caption{Performance (in seconds) comparison of lu using different compilers and autotuning}
\begin{tabular}{c}
  \includegraphics[width=.47\textwidth]{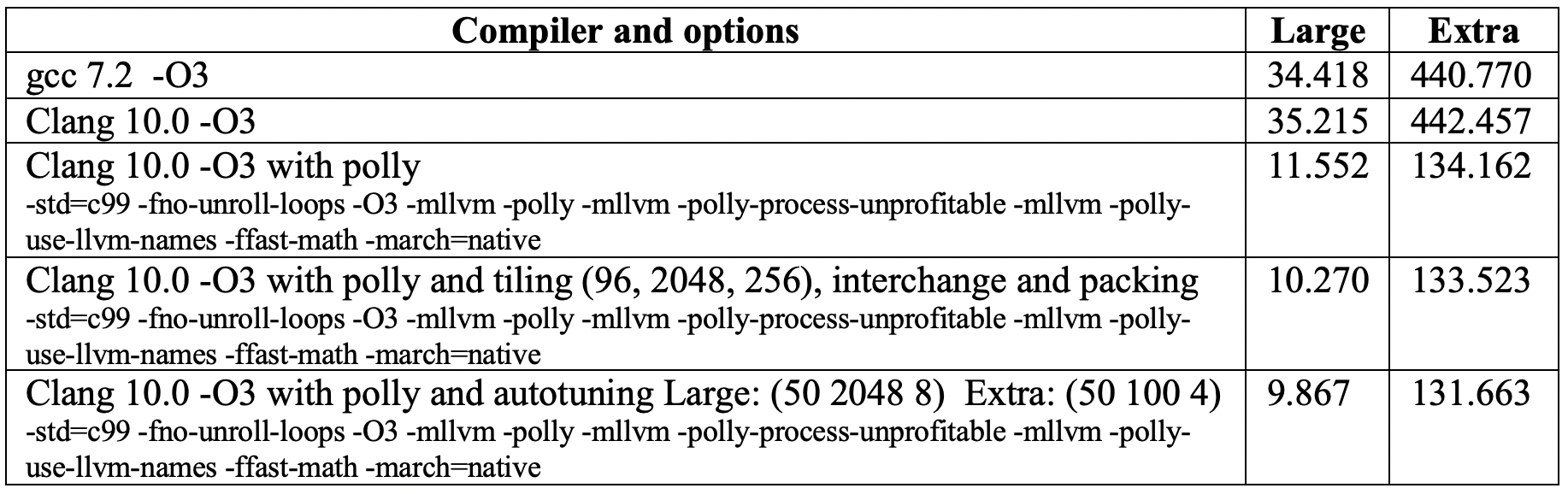}
  \end{tabular}
\label{tab3}       
\end{table} 

Table \ref{tab3} shows the performance comparison of lu using different compilers and autotuning. We find that the best configuration in 200 evaluations has the tile size (50 2048 8) for the large dataset and the tile size (50 100 4) for the extra large dataset using GBRT. We observe that autotuning outperforms the other methods to provide the smallest execution time for both datasets. 

\subsection{Case Study: heat-3d with multiple loop transformations} 

\begin{figure}
\center
 \includegraphics[width=.45\textwidth]{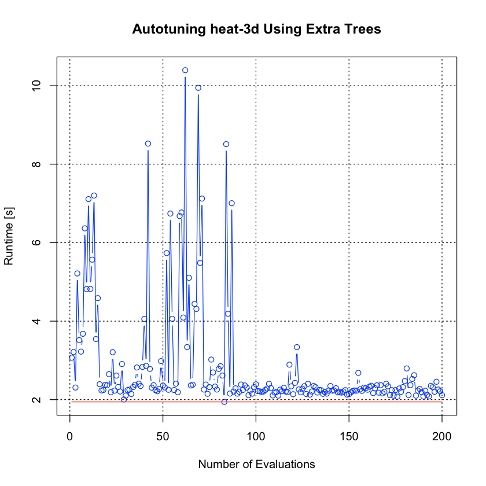}
 \caption{Autotuning heat-3d using ET in 200 evaluations}
\label{fig9}
\end{figure}

We apply multiple loop transformations such as tiling, interchange, and array packing pragmas to the benchmark heat-3d for this case study. We define the six pragma parameters to autotune the benchmark. We use four ML methods---RF, ET, GBRT, and GP---to check which one generates the smallest runtime for which configuration in 200 evaluations for heat-3d with large dataset. We find that ET has the smallest runtime of 1.942 s for the configuration (' ', ' ' ,' ', 100, 64, 128) at Evaluation 83 of 200 evaluations, as shown in Figure \ref{fig9}. We then use ET to autotune the benchmark with the extra large dataset.

\begin{table}
\center
\caption{Performance (in seconds) comparison of heat-3d using different compilers and autotuning}
\begin{tabular}{c}
  \includegraphics[width=.47\textwidth]{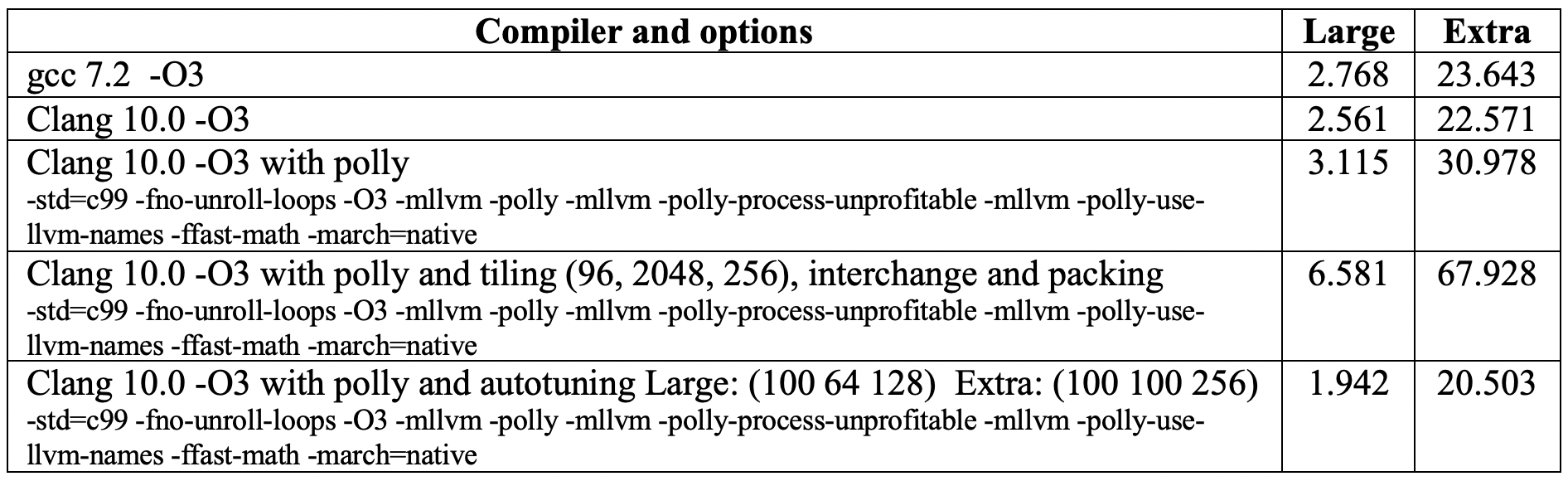}
  \end{tabular}
\label{tab4}       
\end{table} 

Table \ref{tab4} shows the performance comparison of heat-3d using different compilers and autotuning. We find that the best configuration in 200 evaluations has the tile size (100 64 128)  for the large dataset and the tile size (100 100 256) for the extra large dataset using ET. We observe that autotuning outperforms the other methods to provide the smallest execution time for both datasets. 

\subsection{Case Study: covariance with multiple loop transformations} 

\begin{figure}
\center
 \includegraphics[width=.45\textwidth]{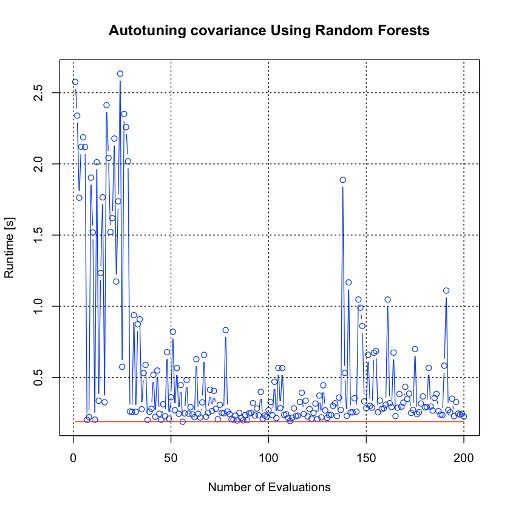}
 \caption{Autotuning covariance using RF in 200 evaluations}
\label{fig10}
\end{figure}

We apply multiple loop transformations such as tiling, interchange, and array packing pragmas to the benchmark covariance for this case study. We define the five pragma parameters to autotune the benchmark. We use four ML methods---RF, ET, GBRT, and GP---to check which one generates the smallest runtime for which configuration in 200 evaluations for covariance with large dataset. We  find that RF has the smallest runtime of 0.188s for the configuration (' ' ,' ', 96, 100, 8 ) at Evaluation 56 of 200 evaluations, as shown  in Figure \ref{fig10}. We then use RF to autotune the benchmark with the extra large dataset.

\begin{table}
\center
\caption{Performance (seconds) comparison of covariance using different compilers and autotuning}
\begin{tabular}{c}
  \includegraphics[width=.47\textwidth]{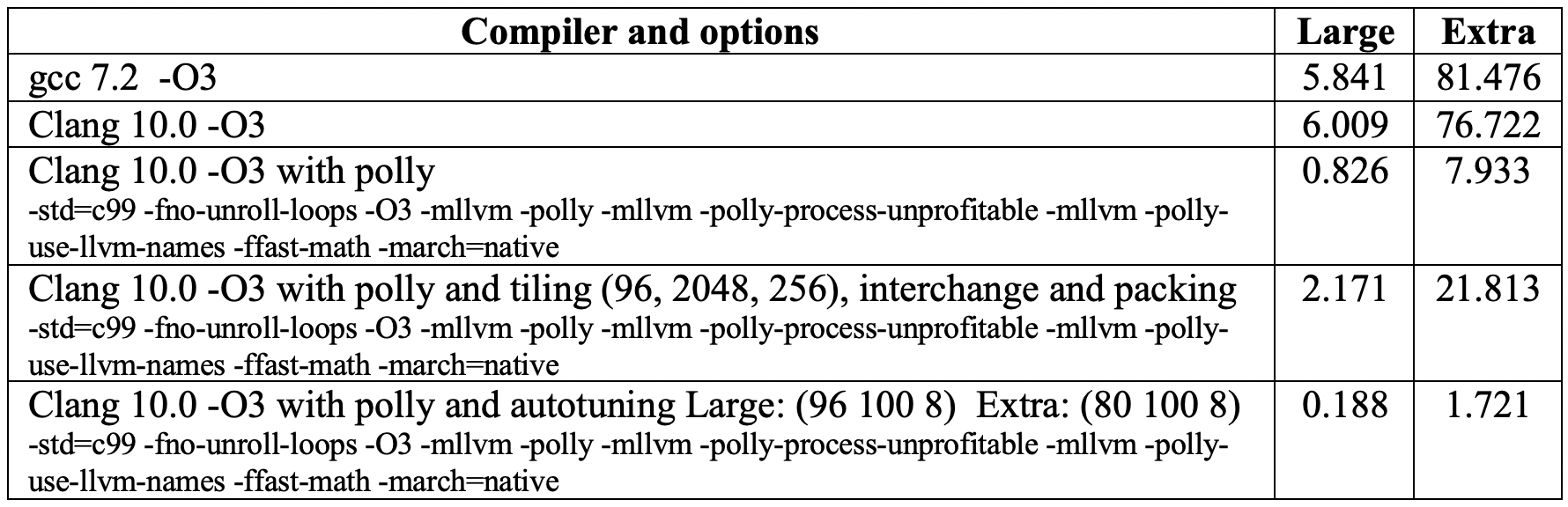}
  \end{tabular}
\label{tab5}       
\end{table} 

Table \ref{tab5} shows the performance comparison of covariance using different compilers and autotuning. We find that the best configuration in 200 evaluations has the tile size (96 100 8)  for the large dataset and the tile size (80 100 8) for the extra large dataset using RF. We observe that autotuning outperforms the other methods to provide the smallest execution time for both datasets. 

\subsection{Case Study: Floyd-Warshall with multiple loop transformations}

We apply multiple loop transformations such as tiling, interchange, and array packing pragmas to the benchmark Floyd-Warshall for this case study. We find that when we compiled the code, the following warning occurred: "floyd-warshall.c:89:5: warning: loop(s) not tiled: transformation would violate dependencies [-Wpass-failed=polly-opt-isl]." That is, the pragmas were ineffective, and Polly applied its default transformation. When we ran the benchmark with the large dataset, it takes more than 135 s shown in Table \ref{tab6}. We note that Clang with Polly causes the benchmark to run very slowly (almost 9 times slower). We investigate what happened as follows.

\begin{table}
\center
\caption{Performance comparison of Floyd-Warshall using different compilers and options}
\begin{tabular}{c}
  \includegraphics[width=.47\textwidth]{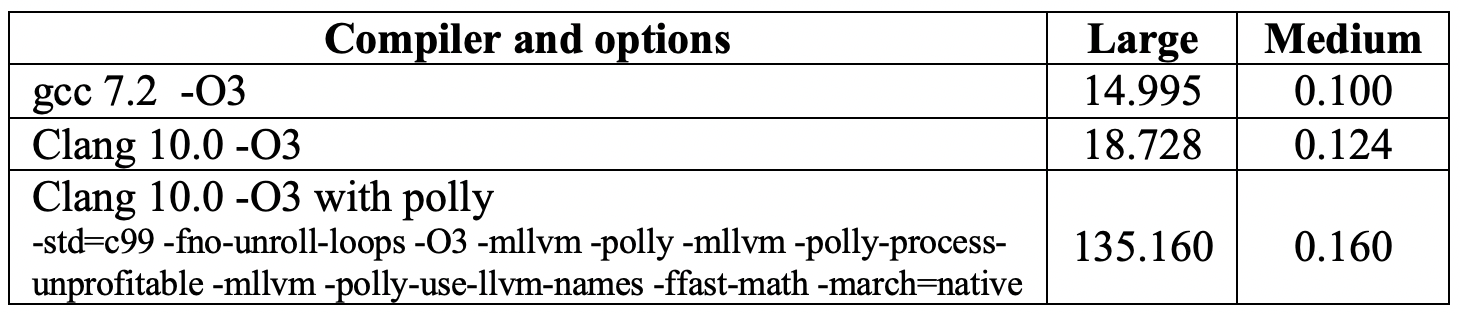}
  \end{tabular}
\label{tab6}       
\end{table}

In the PolyBench Floyd-Warshall code, the main kernel looks like the following.
{\scriptsize
\begin{verbatim}
for (k = 0; k < _PB_N; k++)
  for(i = 0; i < _PB_N; i++)
    for (j = 0; j < _PB_N; j++)
      path[i][j] = (path[i][j] < path[i][k] + path[k][j]) 
         ? path[i][j]
         : (path[i][k] + path[k][j]);
\end{verbatim}
 }
Most accesses need to have the innermost subscript dependent on the fastest iterating induction variable j, meaning consecutive memory accesses in the innermost loop (i.e., spatial locality) enable effective use of cache lines and prefetching by the CPU.

Using Polly's default loop optimization heuristic implemented by ISL (Integer Set Library), the following schedule is applied.

{\scriptsize
\begin{verbatim}
for (c0 = 0; c0 <= 2799; ++c0)
  for (c1 = 0; c1 <= 5598; ++c1)
    for (c2 = max(0, c1 - 2799); c2 <= min(2799, c1); ++c2)
      Stmt_for_body6_i(c0, c2, c1 - c2);
\end{verbatim}
 }

This corresponds to something like the following loop.
{\scriptsize
\begin{verbatim}
for (j = 0; j < _PB_N; j++)
  for (k = 0; k < _PB_N; k++)
    for(i = _PB_N-1; i >= 0; i--)
      path[i][j] = (path[i][j] < path[i][k] + path[k][j]) 
         ? path[i][j]
         : (path[i][k] + path[k][j]);
\end{verbatim}
 }
 
In this variant, the fastest-running index i is not the innermost subscript of any of the array accesses. In other words, all the accesses are strided in memory and will access different cache lines from those of the previous iterations. This is bad for performance.
The default loop nest optimization strategy only considers temporal reuse, but not spatial reuse. Therefore, it does not prioritize keeping the j-loop as the innermost loop with the fastest-running index, making the execution slower than the original loop nest.

\begin{table}
\center
\caption{Performance improvement of Floyd-Warshall using Clang/Polly with additional options and autotuning}
\begin{tabular}{c}
  \includegraphics[width=.47\textwidth]{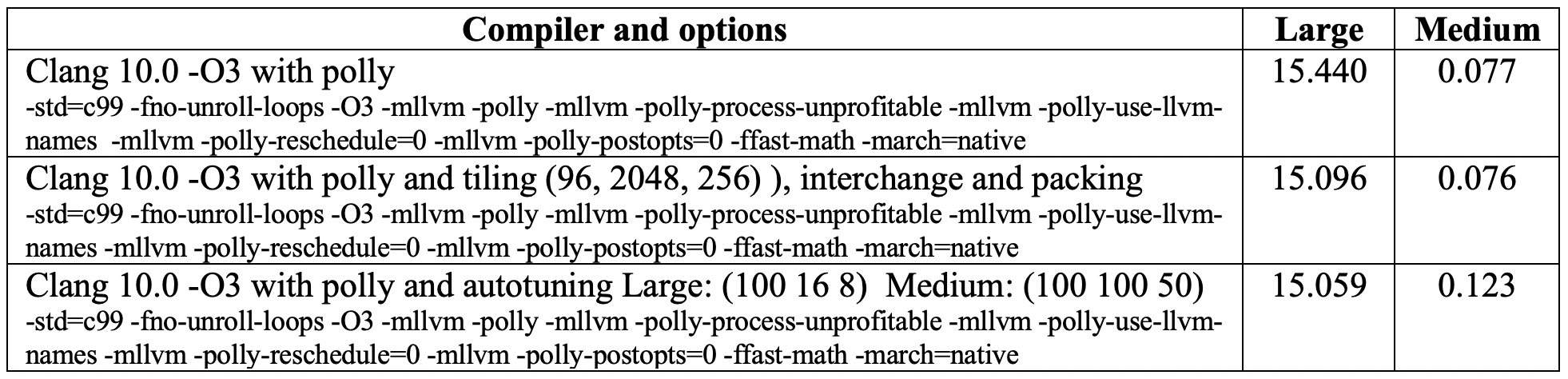}
  \end{tabular}
\label{tab7}       
\end{table} 

\begin{figure}
\center
 \includegraphics[width=.45\textwidth]{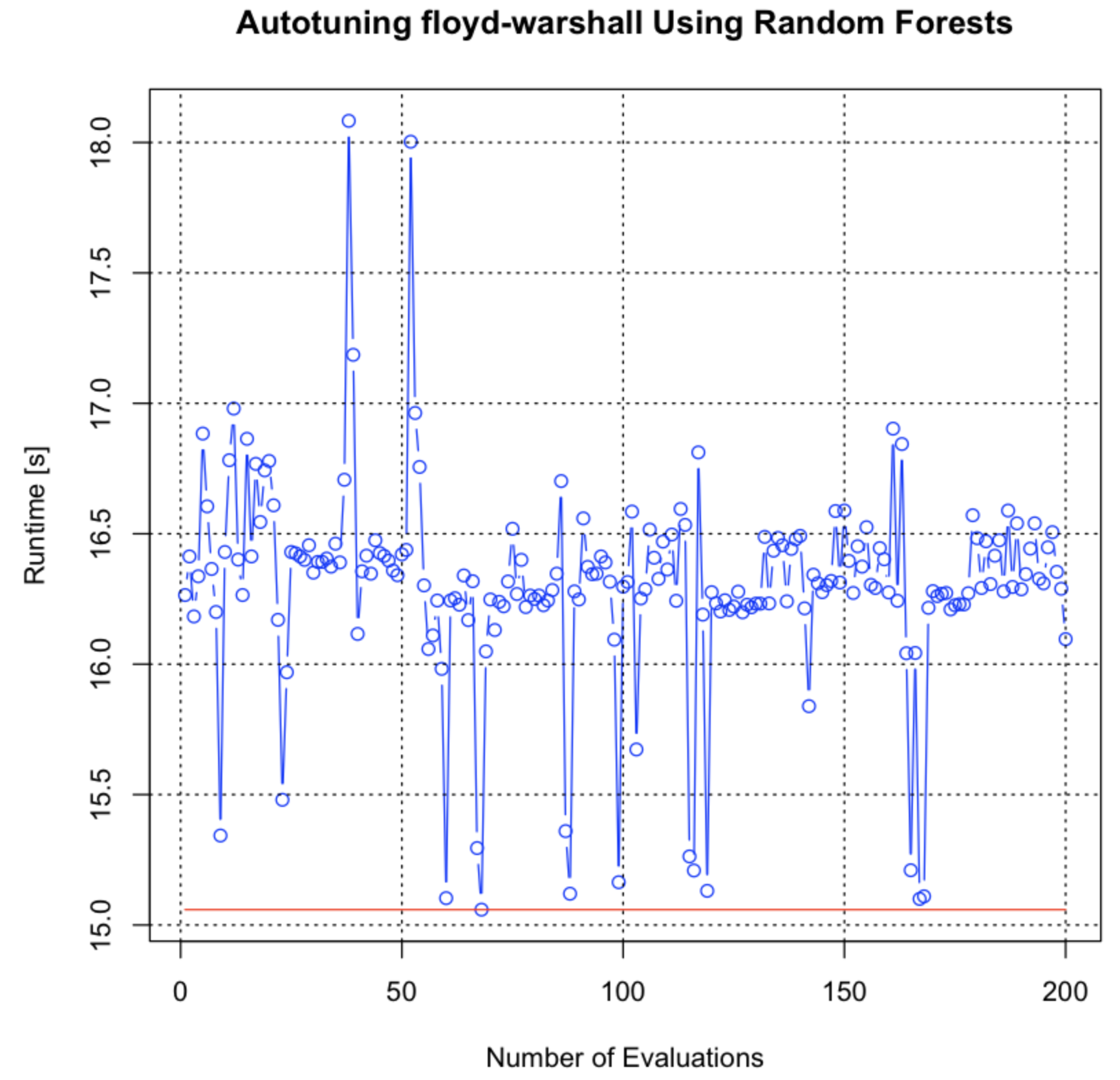}
 \caption{Autotuning Floyd-Warshall using RF in 200 evaluations}
\label{fig12}
\end{figure}

To cope with this situation, we updated \url{https://github.com/SOLLVE/llvm-project/tree/pragma-clang-loop} to include new flags: the flags -mllvm -polly-reschedule=0 -mllvm -polly-postopts=0. These flags make Polly do nothing if no pragma is applied, instead of using the default optimizer. Additionally, and the flag -mllvm -polly-pragma-ignore-depcheck makes Polly apply a transformation even if it cannot confirm that it is semantically correct. 
In this case it is necessary because the ternary operation applies a max-reduction, which is commutative but the computer cannot detect.
Hence, adding the flag forces Polly to apply tiling even though the compiler cannot ensure its semantic legality.
We find that RF has the smallest runtime of 15.059 s for the configuration (' ' ,' ', 100, 16, 8 ) at Evaluation 68 of 200 evaluations for the large dataset in Figure \ref{fig12}. We then use RF to autotune the benchmark with the medium dataset. Compared to Table \ref{tab6}, we observe that the significant performance improvement occurs from around 135s to around 15s in Table \ref{tab7} with Polly by adding these flags -mllvm -polly-reschedule=0 -mllvm -polly-postopts=0. Overall, this benchmark illustrates that a heuristic-based optimization can also regress a program's performance due to being unable to model the entire architecture complexity and unavailability of dynamic information such as the actual execution time used by the autotuning approach.

\section{Summary and Future Work}

We developed a ML-based autotuning framework, and applied the newly developed Clang loop optimization pragmas to six complex PolyBench benchmarks (syr2k, 3mm, heat-3d, lu, covariance, and floyd-warshall) to optimize them. We defined the parameters for these pragmas and then used the autotuning framework to optimize the pragma parameters to improve their performance. We evaluated the effectiveness of four different supervised ML methods used as the surrogate model within Bayesian optimization for each benchmark. The autotuning outperformed the other compiling methods to provide the smallest execution time for the benchmarks syr2k, 3mm, heat-3d, lu, and covariance with both large datasets. An exception is the Floyd-Warshall benchmark because Polly uses heuristics to optimize the benchmark to make it run much slower. To cope with this situation, we provide three compiler option solutions to improve the performance. This autotuning framework is open source and is available from the link in \cite{WK20}.

\begin{figure}
\center
 \includegraphics[width=.4\textwidth]{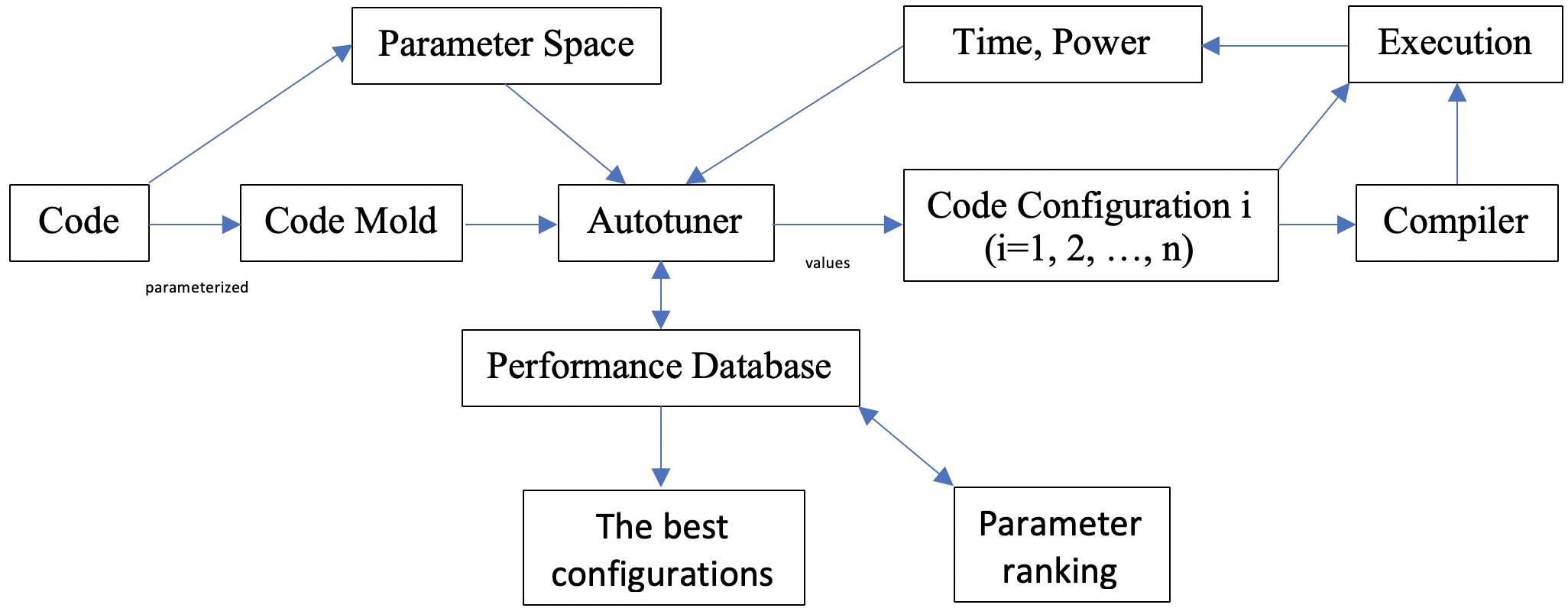}
 \caption{A general autotuning framework}
\label{fig11}
\end{figure}

For future work, based on the performance database, we plan to identify which feature impacts the performance most to aid in the parameter space search.
We will easily extend the current autotuning framework to support various HPC applications because of our symbol representations for the pragmas and related parameters. The current autotuning framework focuses on the application execution time. In \cite{SP18}, autotuning OpenMP codes was investigated for energy efficient HPC systems. In \cite{WM20}, an end-to-end autotuning framework in HPC PowerStack was proposed to tune the power and energy ecosystem. We will extend this autotuning framework to consider power consumption and energy consumption shown in Figure \ref{fig11}. 
For the scripting language applications such as Python, the framework is easily extended to support their direct executions. For the compiler-supported applications such as C, C++, and Fortran with OpenMP and MPI, the framework is able to compile and execute them. When we consider power or energy as the optimal solution, this may change how to do the parameter space search based on the new metric. These parameters can be extended to include application parameters, system environment parameters such as setting number of threads, thread scheduling and affinity, JIT-enabled parameters, power-capping size, and loop transformation parameters, and so on. Future work will also focus on loop autotuning without any human knowledge about the loop pragmas and related parameters. 

\section*{Acknowledgments}
This work was supported in part by LDRD funding from Argonne National Laboratory, provided by the Director, Office of Science, of the U.S. Department of Energy (DoE) under contract DE-AC02-06CH11357, in part by DoE ECP PROTEAS-TUNE, and in part by NSF grant CCF-1801856.

\begin{thebibliography}{}
%
%

\bibitem{AUTO} autotune, https://github.com/ytopt-team/autotune.git.
\bibitem{BG13} P. Balaprakash, R. B Gramacy, and S. M Wild,  Active-Learning-Based Surrogate Models for Empirical Performance Runing, in 2013 IEEE International Conference on Cluster Computing (CLUSTER'13), 2013.
\bibitem{BD18} P. Balaprakash, J. Dongarra, T. Gamblin, M. Hall, J. K. Hollingsworth, B. Norris, and R. Vuduc, Autotuning in High Performance Computing Applications, in Proceedings of the IEEE, 2018. DOI: 10.1109/JPROC.2018.2841200 
\bibitem{BH19} P. Balaprakash, P. Hovland, S. Wild, M. Hall, M. S. Lakhdar, and X. Li, Machine-Learning-Based Performance Modeling and Tuning for High-Performance Computing, Machine Learning for HPC minisymposium in PASC19, June 13, 2019  (talk slides).
\bibitem{BU20} U. Bondhugula, High Performance Code Generation in MLIR: An Early Case Study with  GEMM, arXiv:2003.00532v1 [cs.PF], March 1, 2020. 
\bibitem{CH06} I. Chung and J. K Hollingsworth,  A Case Study Using Automatic Performance Tuning for Large-Scale Scientific Programs, in 15th IEEE International Symposium on High Performance Distributed Computing (HPDC'06), 2006.
\bibitem{CFS} Configspace, https://automl.github.io/ConfigSpace/master/
\bibitem{FE17} T. L Falch and A. C Elster,  Machine Learning-Based Auto-Tuning for Enhanced Performance Portability of OpenCL Applications. Concurrency and Computation: Practice and Experience 29, 8, 2017. 
\bibitem{GO10} M. Gerndt and M. Ott., Automatic Performance Analysis with Periscope, Concurrency and Computation: Practice and Experience 22, 6, 2010. 
\bibitem{KC14} J. Katarzynski and M. Cytowski, Towards Autotuning of OpenMP Applications on Multicore Architectures. http://arxiv.org/abs/1401.4063, 2014.
\bibitem{KF19} M. Kruse and H. Finkel, User-Directed Loop-Transformations in Clang, in International Workshop on OpenMP (IWOMP'19),  Auckland, New Zealand,  September 2019.
\bibitem{LLVM} LLVM compiler infrastructure, https:/llvm.org
\bibitem{LI16} T. M. Low, F. D. Igual, T. M. Smith, and E. S. Quintana-Orti, Analytical Modeling Is Enough for High-Performance BLIS, ACM Transactions on Mathematical Software, Vol. 43, No. 2, Article 12, August 2016.
\bibitem{MA17} A. Marathe, R. Anirudh, N. Jain, A. Bhatele, J. Thiagarajan, B. Kailkhura, J. Yeom, B. Rountree, and T. Gamblin, Performance Modeling under Resource Constraints Using Deep Transfer Learning, in SC17, Nov. 2017.
\bibitem{MS14} S. Muralidharan, M. Shantharam, M. Hall, M. Garland, and B. Catanzaro,  Nitro: A Framework for Adaptive Code Variant Tuning, 2014 IEEE 28th International Parallel and Distributed Processing Symposium (IPDPS'14), 2014. 
\bibitem{MA11} D. Mustafa, A. Aurangzeb, and R. Eigenmann, Performance Analysis and Tuning of Automatically Parallelized OpenMP Applications, in 7th International Conference on OpenMP in the Petascale Era (IWOMP'11), 2011.
\bibitem{OP17} W. F Ogilvie, P. Petoumenos, Z. Wang, and H. Leather, Minimizing the Cost of Iterative Compilation with Active Learning,  2017 International Symposium on Code Generation and Optimization, 2017. 
\bibitem{POLL} Polly, a High-Level Loop and Data-Locality Optimizer and Optimization Infrastructure for LLVM, https://polly.llvm.org.
\bibitem{RB16} A. Roy, P. Balaprakash, P.l D Hovland, and S. M Wild, Exploiting Performance Portability in Search Algorithms for Autotuning, in 2016 IEEE International Parallel and Distributed Processing Symposium Workshops: iWAPT, 2016. 
\bibitem{SCIO} scikit-optimize, https://github.com/pbalapra/scikit-optimize.git.
\bibitem{SP18} Silvano, C., Palermo, G., Agosta, G., Ashouri, A.H., Gadioli, D., Cherubin, S., Vi- tali, E., Benini, L., Bartolini, A., Cesarini, D., Cardoso, J., Bispo, J., Pinto, P., Nobre, R., Rohou, E., Besnard, L., Lasri, I., Sanna, N., Cavazzoni, C., Cmar, R., Marti- novic, J., Slaninova, K., Golasowski, M., Beccari, A.R., Manelfi, C.,  Autotuning and Adaptivity in Energy Efficient HPC Systems: The ANTAREX toolbox, in  15th ACM International Conference on Computing Frontiers( CF'18), 2018. 
\bibitem{SL12} J. Snoek, H. Larochelle, and R. P. Adams, Practical Bayesian Optimization of Machine Learning Algorithms, Advances in Neural Information Processing Systems 25 (NIPS 2012), 2012.
\bibitem{SOLL} SOLLVE LLVM, https://github.com/SOLLVE/llvm-project.git.
\bibitem{SJ19} V. Sreenivasan, R. Javali, M. Hall, P. Balaprakash, T. R. W. Scogland, B. R. de Supinski, A Framework for Enabling OpenMP Autotuning, in International Workshop on OpenMP (IWOMP'19), Auckland, New Zealand,  Sept. 2019.
\bibitem{TC02} C. Tapus, I. Chung, and J. K. Hollingsworth,  Active Harmony: Towards Automated Performance Tuning,  the 2002 ACM/IEEE Conference on Supercomputing (SC02), 2002. 
\bibitem{TH11} A. Tiwari and J. K Hollingsworth,  Online Adaptive Code Generation and Tuning,  in 2011 IEEE International Parallel \& Distributed Processing Symposium (IPDPS'11),  2011.
\bibitem{TC09} A. Tiwari, C. Chen, J. Chame, M. Hall, and J. K Hollingsworth, A Scalable Auto-tuning Framework for Compiler Optimization, in 2009 IEEE International Symposium on Parallel \& Distributed Processing( IPDPS'09). 2009. 
\bibitem{TN18} J.J. Thiagarajan, N. Jain, R. Anirudh, A. Gimenez, R. Sridhar, A. Marathe, T. Wang, M. Emani, A. Bhatele, and T. Gamblin, Bootstrapping Parameter Space Exploration for Fast Tuning, in ICS'18, June 2018.
\bibitem{WM20} X. Wu, A. Marathe, S. Jana, O. Vysocky, J. John, A. Bartolini, L. Riha, M. Gerndt, V. Taylor and S. Bhalachandra, Toward an End-to-End Auto-Tuning Framework in HPC PowerStack, in  Energy Efficient HPC State of Practice 2020, Kobe, Japan, Sept. 14, 2020.
\bibitem{WK20} X. Wu, M. Kruse, P. Balaprakash, B. Videau, H. Finkel, and P. Hovland, Autotuning Polybench Benchmarks with Clang Loop Optimization Pragmas, Technical Report, April 16, 2020. https://github.com/ytopt-team/autotune/blob/master/Benchmarks/polybench-autotune.pdf. See these benchmarks from https://github.com/ytopt-team/autotune/blob/ master/Benchmarks. 
\bibitem{YTO} ytopt, https://github.com/ytopt-team, https://github.com/ytopt-team/ ytopt.git.
\bibitem{YP16} T. Yuki and L. Pouchet, PolyBench 4.2, May 9, 2016. https://sourceforge.net/projects/polybench/

\end{thebibliography}
\end{document}